# How the notion of ACCESS guides the organization of a European research infrastructure: the example of DARIAH

1. What is DARIAH ?

DARIAH (Digital Research Infrastructure for the Arts and Humanities) is an infrastructure developed under the auspices of the European Commission, it aims to organize communities in those fields, to develop interdisciplinary projects, promoting in particular the digital dimension of humanities and arts research by disseminating good practices, and providing tools and services. In legal terms, it is what is called an ERIC, that is to say a European Research Infrastructure Consortium.

DARIAH had already a long existence as an unofficial structure since 2005, but it becomes fully established as an ERIC in 2014. It brings together 17 countries in Europe which makes it the biggest ERIC in terms of members, but also in terms of scientific fields it covers.

2. Why access supports the creation of the infra ?

Many of our pre-existing understandings of how access impacts upon humanities and arts research date from the days of the library collection, and it is common to limit the notion of access to research data. But within the framework of a research infrastructure consortium, the notion of access is made more complex. We will evidence this different paradigm here with five examples, which we will develop in turn by explaining both the constraints and the solutions that are envisaged.

According to the Cambridge Dictionary, "access" has two kinds of meanings: the first one concerns "the method or possibility of getting near a place or a person" and the second one "the right or opportunity to use or look at something". Both meanings are interesting in terms of creating an expanded understanding of access in field of Digital Humanities. Even if we mostly think about the second one (open access, open data, and so on), the first one shows the necessity of being near our community and highlights the fact that thanks to digital tools we are nearer and nearer despite the distance and we are able to work together being in different places.

5 points have been raised to show how access plays a strong role :
_ Managing interdisciplinarity
_ Managing tensions between national and international perspectives
_ Speaking to whom?
_ Managing tools
_ Building collaborative tools

1. **The Strategic Action Plan document : Build upon access notion**

I will take a specific example which explain how DARIAH works and how we try to solve those problems : the redaction and the content of a Strategic Action Plan document, the « 25 key Actions for a stronger DARIAH by 2020 » :

This document builds upon the conclusions and syntheses of a large amount of published research working papers and internal project documents. Also key to its genesis has been the creation of a number of internal planning documents, largely at the end of 2016-2017, capturing key discussions, exercises (like a thorough SWOT analysis), and experiences of the period directly leading to the paper's development. Finally, the completed document is the result of a broad and systematic programme of consultation across the community (as you can see on the screen). This shows what access means inside DARIAH and how we try to share informations within all the community. This methodology is a first step toward access.

- Managing interdisciplinarity

The first question needs to understand the same thing under the expression « Arts and Humanities ». Which disciplines are concerned ? And, how can we work together ? This implies to know who do what and where, to have common tools which can be used by researchers in different fields and different countries.
DARIAH is working on that by :
_ creating a strong network by its registry (still in progress, but planned to be finished in October 2017) **[examples are taken from the future new website which will be online in September…]**
_ its organization in WG, which are specific groups of interests characterized by their interdisciplinarity. Transversal WGs + actions listed in the point 6 of the document (Support and promote research communities through the DARIAH Working Groups and other structures, including regional hubs) show how interdisciplinary question is also a question of access.

- Managing tensions between national and international perspectives

Access also has a political dimension. When the same tool is developed in parallel in two different countries, how can we know how to assign credit for the development? Which one (and why) should be valued by DARIAH? To resolve this tension, the DARIAH community is coordinated around National Representatives who are engaged in complementary, rather than competitive, work. This work is based in particular on the dissemination of information about the activities of the national teams and the projects in which they are involved.

Moreover, some teams may wish to retain the rights or control of their tools and may not wish to make them accessible to other communities without compensation. The selection of tools which were developed at the same time in different countries is made regarding their international use and thanks to several exchanges. This is in particular the aim of the Marketplace (point 2).

The strong structuration of DARIAH helps to solve tensions which can emerge : a JRC and a NCC : both committees form the SMT : an instance which aims to develop DARIAH strategy with both a national point of view and an international strategy of development.

- Speaking to whom?

This is, actually, the question of the definition of communities. Which are they ? Defined by the covered field (which is, as we've seen, very broad), but not only : researchers, students, already used in DH tools or not ? private companies ? This is very vague. And even the place « Europe », could be problematic. Is is geographical, political or economical boudary ?
Speaking about access imply to define limits or to explain why there could be no limit (if this option could really exist).
Imperatives toward democratizing the benefits DARIAH can bring come at such junctures into conflict with the possibility that too much access could dilute the infrastructure's effectiveness, distort its scale or divert its mission. Again, this is an aspect related to the question of access to which the infrastructure must respond and on which we shall give a few quick lines of reflection.
This specific question of access is partially solved with different statutes of partner/memberships + different kind of projects.

Regarding the people, it has been decided to include mainly students which have research projects (master + PhD students) and researchers + teachers. This is for instance to contribute to develop the field of Digital Humanities.

Finally, the third point of the document about the organization of external communications (9 actions have been raised to specific communities) gives also some answers to that point.

- Managing tools

This point is particularly well recognised within the DH community, since it questions the interoperability of tools. For DARIAH, questions regarding managing access to tools arises in terms of languages, content, formats and, of course, sustainability. Specific attention is therefore paid to this aspect and in particular to data hosting. Interoperability question finds some answer both within the Marketplace's creation and thanks to WGs : by their projects, they can highlight the needs in terms of tools and relations between them.

This perspective on access reflects the importance of the trust that must be established between the partner countries, in particular with regard to intellectual property, as suggested in the next and final point.

- Building collaborative tools

As a digital infrastructure, DARIAH brings together 17 countries and a range of diverse disciplinary communities. In this sense, it involves collaborative work that relies mainly on the use of digital tools. But one problem remains: the too easy access to collaborative tools developed by companies that do not share the same conception of intellectual property and data security. Tools such as Google doc and Google drive, etc. are unavoidable in the context of collaboration between researchers, but the access in this case is so easy that their existence prevents the development of alternative tools that correspond more closely to the specificities of scientific exchanges. And more, it is not acceptable regarding the Open Access and the respect of intellectual property, etc. DARIAH tries to organizes itself, as well as the scientific digital practices, with the Marketplace for instance but also by promoting specific tools like a wiki, developed by Dariah Germany or by using a Basecamp project, which is not in Open Access but is offered to DARIAH members for collaborative projects. DARIAH relies on national skills such as French one's for hosting.

Conclusion :

As a conclusion, let's say that DARIAH is a very young European infrastructure with such great challenges. Most of them are controlled or achieved thanks to its specific structuration, its Adaptation to the community, its distributed communication, its Innovation by reuse and transforming the existing and finally its focus on training. DARIAH supports both technical developments to help researchers and scientific programs. It defends strongly Open Access and tries to show it in each of its development. Access, for DARIAH, means as well introducing people to new practices as gather people.

**Suzanne Dumouchel,**

**DH Montréal, 10/08/2017**